\documentclass[pra,aps,showpacs,twocolumn]{revtex4}
\usepackage{epsfig}
\date{ }

\begin{document} 
\title{Dynamics and kinetics of quasiparticle decay in a nearly-one-dimensional 
degenerate Bose gas} 
\author{I. E. Mazets} 
\affiliation{Vienna Center for Quantum Science and Technology, Atominstitut, TU Wien, 
1020 Vienna, Austria;   \\ 
Ioffe Physico-Technical Institute, 194021 St.Petersburg, Russia} 

\begin{abstract} 
We consider decay of a quasiparticle in a nearly-one-dimensional quasicondensate of trapped atoms, 
where virtual excitations of transverse modes break down one-dimensionality and integrability, 
giving rise to effective three-body elastic collisions. We calculate the matrix element for the 
process that involves one incoming quasiparticle and three outgoing quasiparticles. Scattering 
that involves low-frequency modes with high thermal population results in a diffusive dynamics 
of a bunch of quasiparticles created in the system. 
\end{abstract} 

\maketitle 

\section{Introduction} 
\label{sec:1} 
An uniform one-dimensional (1D) system of indistinguishable bosons interacting with each other 
via pairwise delta-functional potential is known to be integrable and described by the Lieb-Liniger 
model \cite{LL}. In an integrable system, the number of integrals of motion equals to the 
number of degrees of freedom. Such an equality, on the one hand, facilitates analytical treatment 
of integrable models and, on the other hand, make their dynamics quite distinct from the dynamics 
of non-integrable many-body systems. In the course of its evolution, an integrable system 
always ``remembers" the information on its initial conditions, and relaxation to a thermal equilibrium  
state does not occur. A general review of integrable models can be found, e.g., in Ref. \cite{Thacker}. 
The Lieb-Liniger model can be implemented in an ultracold-atom experiment  
in optical lattices \cite{je1} or on atom chips \cite{je2}. The conditions of the 1D regime require  
both the temperature and mean interaction energy per atom being well below the excitation quantum 
$\hbar \omega _\perp $ of the radial trapping (harmonic oscillator) Hamiltonian: 
\begin{equation} 
T\lesssim \hbar \omega _\perp ,  \qquad  g_\mathrm{1D} n_\mathrm{1D} \lesssim \hbar \omega _\perp . 
\label{w.1} 
\end{equation} 
Here temperature $T$ is measured in units of energy (i.e., we set $k_\mathrm{B}=1$) and $n_\mathrm{1D}$ 
is the linear density of atoms. Since the effective 1D interaction strength for atoms under radial 
harmonic confinement  is (far below the confinement-induced resonance \cite{Olsh}) 
$g_\mathrm{1D}=2\hbar \omega _\perp a_s$, where $a_s$ is the atomic $s$-wave scattering length 
Under these conditions \cite{explsm}, the latter of two Eqs. (\ref{w.1}) reads 
\begin{equation} 
n_\mathrm{1D}a_s \lesssim 1.    \label{w.2} 
\end{equation} 
If Eqs. (\ref{w.1}) hold then 
radial motion of atoms is confined to the ground state of the radial trapping Hamiltonian. 
Ultracold atomic systems 
in the 1D regime can be prepared in optical lattices \cite{WeissNC} and on atom chips \cite{Ho1}. 
However, in reality no system is perfectly 1D, but the actual question is, on which timescale it can be 
described as 1D. Since the thermal population of the excited states is strongly suppressed by 
the exponential Boltzmannian factor, already $T\approx 0.2\, \hbar \omega _\perp $ \cite{Ho1} corresponds to a deeply  
1D regime in terms of thermal excitations. On the other hand,  the influence of atomic interactions is a more 
complicated and interesting issue. For ultracold atoms under tight lateral confinement one-dimensionality and, 
hence, integrability are lifted by atomic interactions causing virtual population of excited 
radial modes. The role of the virtual radial excitations in 
the dynamics of ultracold atomic gases in tight waveguides has been first studied in the context of 
macroscopic flow of degenerate atomic gas through a waveguide \cite{SPR} and decay \cite{Muryshev} or 
inelastic collisions \cite{Malomed}  
of mean-field solitons. On the microscopic level, a second-order perturbative collisional process with a  
radially excited virtual (intermediate) state give rises to effective three-body elastic scattering that 
has been suggested as the source of thermalization 
in ultracold atomic gases on atom chips \cite{Mazets1,Mazets2}.  

\begin{figure}[h]
\vspace*{9mm}
\begin{center}
\epsfig{file=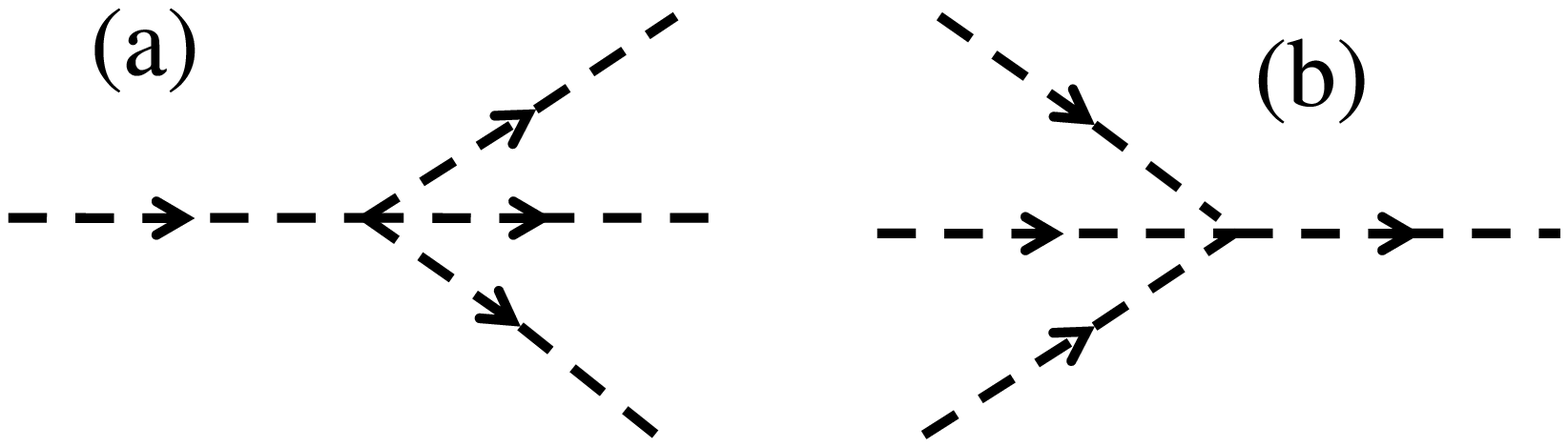,width=65mm} 
\end{center} 
\begin{caption}
{Feynman diagrams of the processes studied in the present paper: (a) splitting of an elementary 
excitation into three excitations, (b) the reciprocal process. 
\label{F1} }
\end{caption}
\end{figure} 

Damping of a fast particle motion in a quasi-1D bosonic system due to the effective three-body 
process shown in Fig.~\ref{F1} has been studied recently \cite{Glazman}, and the contribution 
of events with small momentum transfer to the damping rate has been found significant in a certain 
parameter range. The reason is the bosonic amplification of scattering into modes with high 
thermal population \cite{Glazman}. In the present paper we derive 
(i) the vertex for the diagrams in Fig.~\ref{F1} and (ii) a kinetic equation describing damping 
relaxation of a quasiparticle in a degenerate nearly-1D bosonic system and its diffusion-type 
limit describing the elementary excittation dynamics induced by collisions with a small momentum transfer.  

Using our theoretical estimations, we conclude that the three-body elastic processes responsible for 
the integrability breakdown can be detected experimentally by observing the damping  
dynamics of an ensemble of fast particles on relatively short time scales. 

\section{Physical  model} 
\label{sec:2}
The Hamiltonian of a 1D system of identical bosons with an additional term accounting for effective 
three-body interactions via virtual radial excitations reads 
\begin{equation} 
\hat{H} =\hat{H}_\mathrm{LL}+\hat{H}_\mathrm{ 3b } , 
\label{w.3} 
\end{equation}  
where $\hat{H}_\mathrm{LL}$ is the Hamiltonian of the Lieb-Liniger model written 
in terms of the 1D number density $\hat{n}$ and phase $\hat{\phi }$ operators \cite{Popov,MC} 
\begin{equation} 
\hat{H}_\mathrm{LL}=\int _0^Ldz\, {\bigg [} \frac {\hbar ^2}{2m} \frac {\partial \hat{\phi }}{\partial z}
\hat{n} \frac {\partial \hat{\phi }}{\partial z} +\frac {\hbar ^2}{8m} 
\hat{n}^{-1} \left( \frac {\partial \hat{n }}{\partial z}\right) ^2 +\frac {g_\mathrm{1D}}2\hat{n}^2{\bigg ]},
\label{w.4}
\end{equation} 
where $m$ is the atomic mass and $L$ is the quantization length. 
The commutation rule for the phase and density operators is 
$ [ \phi (z), n(z^\prime ) ]=-i\delta (z-z^\prime ) $. 
The three-body collisional dynamics is taken into account by the term 
\begin{equation} 
\hat{H}_\mathrm {3b}=-\frac \xi 2 \hbar \omega _\perp a_s^2 \int _0^Ldz\, \hat{n}^3 , 
\label{w.5} 
\end{equation} 
where $\xi =4 \ln \frac 43 $. Note that an error in the prefactor of $\hat{H}_\mathrm {ni}$ \cite{Mazets1} 
has been corrected later \cite{Mazets2,Glazman}. Of course, Eq. (\ref{w.5}) is only a lowest term 
in expansion of the integrability-breaking interaction in powers of the small parameter $n_\mathrm{1D}a_s$. 
The full effective interaction due to excitation of radial modes contains cubic, quartic, quintic in 
$\hat{n}$ terms and supports a stable ground state. This can be seen from the analogy with the 
mean-field variational approach \cite{SPR,Mazets2}. However, if Eq. (\ref{w.2}) holds then 
Eq. (\ref{w.5}) is sufficient for perturbative calculation of the rates of the processes shown 
schematically in Fig.~\ref{F1}.  

From now on, we scale energy to $g_\mathrm{1D} n_\mathrm{1D}$ and set the healing length 
$\zeta _\mathrm{h}=\hbar /\sqrt{m g_\mathrm{1D} n_\mathrm{1D}}$
as the  length unit, i.e., $\hat{H}\equiv  g_\mathrm{1D} n_\mathrm{1D} \hat{\bar{H}}$ and 
$z=\zeta _\mathrm{h} \bar{z}$. In what follows, we omit, for the sake of 
compactness of notation, the bar over scaled values. It will not lead to confusion, since we will 
mention explicitly the return to usual units. In the new, dimensionless form Eqs. (\ref{w.4},~\ref{w.5}) 
read simply as 
\begin{equation} 
\hat{H}_\mathrm{LL}=\int _0^Ldz\, {\bigg [} \frac {1}{2} \frac {\partial \hat{\phi }}{\partial z}
\hat{n} \frac {\partial \hat{\phi }}{\partial z} +\frac {1}{8} 
\hat{n}^{-1} \left( \frac {\partial \hat{n }}{\partial z}\right) ^2 +\frac 1{2n_\mathrm{1D}}
\hat{n}^2{\bigg ]},
\label{w.4a}
\end{equation}
\begin{equation} 
\hat{H}_\mathrm {ni}=-\frac {\kappa _\mathrm{3b}}{n_\mathrm{1D}^2} \int _0^Ldz\, \hat{n}^3 , 
\label{w.5a} 
\end{equation}
where $\kappa _\mathrm{3b}= \frac 14 \xi  n_\mathrm{1D} a_s$. 

We expand the phase and density operators in plane waves. The average 1D density $n_\mathrm{1D}=N/L$ is an 
integral of motion has a well-defined value since the total number $N$ of atoms is fixed. The expansions 
read as 
\begin{equation} 
\hat{n}=n_\mathrm{1D}+\sum _k \hat{n}_k\frac {e^{ikz}}{\sqrt{L}}, \quad 
\hat{\phi }=\Phi _0+\sum _k \hat{\phi }_k\frac {e^{ikz}}{\sqrt{L}}. 
\label{w.6}
\end{equation} 
The global phase $\Phi _0$, subject to phase diffusion even if $N=$ const \cite{Lewenstein}, will not appear 
in further expressions. The sums in Eq. (\ref{w.6}) are taken of non-zero (both positive and negative) 
integer multiples of $2\pi /L$. The commutation rule for the phase and density operators in the momentum 
representation are 
\begin{equation} 
[ \hat{\phi }_k, \hat{n}_{-k^\prime }] = -i\delta _{kk^\prime }. 
\label{w.7} 
\end{equation} 

By re-grouping terms we express the full Hamiltonian of the system as $\hat{H} = E_\mathrm{gr} +
\hat{H}_\mathrm{h}+\hat{H}_\mathrm{a}+\hat{H}_\mathrm{ni}$, where $E_\mathrm{gr}$ is the ground state 
energy, the two next terms represent the harmonic 
\begin{equation} 
\hat{H}_\mathrm{h}=\sum _k \left[ \frac {n_\mathrm{1D}k^2}2\hat{\phi }_k\hat{\phi }_{-k} +
\left( \frac {k^2}{8n_\mathrm{1D}} +\frac {G_\mathrm{rn}}{2n_\mathrm{1D}}\right) 
\hat{n}_k\hat{n}_{-k} \right] 
\label{w.8} 
\end{equation}  
and anharmonic 
\begin{equation} 
\hat{H}_\mathrm{a}=\frac 1{\sqrt{L}}\sum _{q,q^\prime } \frac {qq^\prime }2\left( - \hat{\phi }_q 
\hat{n}_{-q-q^\prime }\hat{\phi }_{q^\prime } +\frac 1{4n_\mathrm{1D}^2} \hat{\nu }_{-q-q^\prime } 
\hat{n }_q\hat{n }_{q^\prime } \right)  
\label{w.9}  
\end{equation}  
parts of the Lieb-Liniger model Hamiltonian.  Operator $\hat{\nu }_k$ in  Eq. (\ref{w.9}) is 
defined as 
\begin{eqnarray} 
\hat{\nu }_k&=&\frac {n_\mathrm{1D}}{\sqrt{L}} \int _0^L dz\, e^{-ikz} \left( n_\mathrm{1D}^{-1} -\hat{n}^{-1}\right) 
\nonumber \\ & = & 
\hat{n}_k-\frac 1{\sqrt{L} n_\mathrm{1D}} \sum _{k^\prime } \hat{n}_{k-k^\prime }  
\hat{n}_{k^\prime }+\dots ~. 
\label{w.10} 
\end{eqnarray} 
$G_\mathrm{rn}=1-\frac 34 \xi n_\mathrm{1D} a_s$ in Eq. (\ref{w.8}) is the effective strength of 
pairwise interactions renormalized by the presence of the three-body collisions. In usual units, 
the coupling strength $g_\mathrm{1D}$ renormalizes to $g_\mathrm{1D}^\prime =Gg_\mathrm{1D}$. 
However, since Eq. (\ref{w.2}) holds by assumption of the 1D regime, we use in what follows an 
approximation $G_\mathrm{rn}\approx 1$. 

The three-body interaction term that makes our model non-integrable is 
\begin{equation} 
\hat{H}_\mathrm {ni}=-\frac {\kappa _\mathrm {3b}}{\sqrt{L} n_\mathrm{1D}^2} \sum _{q,q^\prime } 
\hat{n} _{-q-q^\prime }\hat{n} _q \hat{n}_{q^\prime } . 
\label{w.11}  
\end{equation}  

Neglect for a moment the anharmonic part (\ref{w.9}) of the Lieb-Liniger model Hamiltonian. The retained 
harmonic part (\ref{w.8}) is easily diagonalized \cite{Popov}: $\hat{H}_\mathrm{h}=\sum _k \epsilon _k \hat{b}_k ^\dag \hat{b}_k$, 
where Bogoliubov elementary excitation energy is 
\begin{equation} 
\epsilon _k =|k| \sqrt{ 1+k^2/4},            \label{w.11bis} 
\end{equation} 
and $\hat{b}_k$ ($\hat{b}_k ^\dag $) is the 
annihilation (creation, respectively) operator for an excitation with the wavenumber $k$. The density and phase operators are then 
\begin{equation} 
\hat{n}_k=\sqrt{n_\mathrm{1D} S_k}(\hat{b}_k+\hat{b}_{-k} ^\dag ), \qquad \hat{\phi }_k=
\frac {\hat{b}_k-\hat{b}_{-k} ^\dag }{2i\sqrt{n_\mathrm{1D} S_k}},
\label{w.21} 
\end{equation} 
where 
\begin{equation} 
S_k = \frac { k^2}{2\epsilon _k }=\frac {|k|}{\sqrt{4+k^2}} 
\label{w.21bis}   
\end{equation}
is the static structure factor of the quasicondensate at zero temperature.  In this (harmonic) 
approximation the operator $\hat{n}_k$ annihilates one elementary excitation with the wave number $k$ 
(or creates an excitation with the wave number $-k$), and the interaction (\ref{w.11}) can cause only 
conventional Beliaev \cite{Beliaev,Liu} or Landau \cite{Popov,Liu,HM} damping. However, these types of 
relaxation caused by decay of an elementary excitation into two excitations or by the reciprocal process 
are completely suppressed in 1D by energy and momentum conservation, in contrast to the 2D and 3D cases. 
To calculate the rates of processes shown in Fig.~\ref{F1}, we need to diagonalize (in some approximation) 
the whole Lieb-Liniger Hamiltonian $\hat{H}_\mathrm{h} +\hat{H}_\mathrm{a}$. The unitary transformation 
providing this diagonalization transforms $\hat{n}_q$ into an operator containing correction term that is 
nonlinear in creation (annihilation) operators of elementary excitations and thus allows for the 
processes shown in in Fig.~\ref{F1}. The transformation is very similar to the polaronic transformation 
\cite{stat}, but, unlike the latter, does not involve impurity particles. In other words, the idea is to demonstrate 
that an elementary excitation in the Lieb-Liniger model corresponds to a phase-density wave at a certain wavelength, 
dressed by virtual phase-density waves. 

To obtain analytic results, we set $\hat{\nu }_k\approx \hat{n}_k$ and thus obtain 
\begin{equation} 
\hat{H}_\mathrm{a}\approx \frac 1{\sqrt{L}}\sum _{q,q^\prime } \frac {qq^\prime }2\left( - \hat{\phi }_q 
\hat{n}_{-q-q^\prime }\hat{\phi }_{q^\prime } +\frac 1{4n_\mathrm{1D}^2} \hat{n}_{-q-q^\prime } 
\hat{n }_q\hat{n }_{q^\prime } \right)  .
\label{w.12}  
\end{equation}  
Before dealing with the particular Hamiltonian of the problem, we discuss the diagonalization procedure in 
general. 

\subsection{Overview of the diagonalization procedure} 
\label{subsec:2.1} 
Consider a Hamiltonian 
\begin{equation} 
\hat{\cal H}= \hat{\cal H}_0+\varepsilon \hat {\cal W},       \label{w.13} 
\end{equation} 
where the unperturbed Hamiltonian $\hat{\cal H}_0$ is diagonal in the 
orthonormal basis  $\{ |\lambda \rangle \} $:
\begin{equation}
\hat{\cal H}_0=\sum _\lambda E_\lambda ^{(0)} |\lambda \rangle \langle \lambda | .
\label{w.14} 
\end{equation} 
The perturbation operator contains the small parameter $\varepsilon $. More precisely, 
$\varepsilon \langle \lambda ^\prime |\hat {\cal W}|\lambda \rangle \ll \left| E_\lambda ^{(0)}-
E_{\lambda ^\prime } ^{(0)}\right| $. In what follows we assume that the diagonal matrix elements of 
$\hat {\cal W}$ is the basis $\{ |\lambda \rangle \} $ are zero (if not, we can eliminate them by 
including $\varepsilon \langle \lambda  |\hat {\cal W}|\lambda \rangle $ into $E_\lambda ^{(0)}$). 

The basis functions $| \tilde{\lambda } \rangle $ that diagonalize the full Hamiltonian (\ref{w.13}) 
\begin{equation} 
\hat{\cal H}=\sum _\lambda E_\lambda  |\tilde{\lambda }\rangle \langle \tilde{\lambda }|
\label{w.13bis} 
\end{equation} 
are related to the old basis via unitary transformation 
\begin{equation} 
|\tilde{\lambda }\rangle = e^{\hat{R}}  |{\lambda }\rangle , 
\label{w.15} 
\end{equation} 
where the operator $\hat{R}=-\hat{R}^\dag $ is anti-Hermitian. In the pertzurbative approach, $\hat{R}$ 
can be expanded in series in the powers of the small parameter 
\begin{equation} 
\hat{R} =\sum _{j=1}^\infty \varepsilon ^j \hat{R}_j .        \label{w.16} 
\end{equation} 
The standard quantum-mechanical perturbation theory \cite{qmpt} yields explicit expressions for the lowest-order 
terms in the expansion (\ref{w.16}), in particular, 
\begin{equation} 
\hat{R}_1=\sum _{\lambda ^\prime \neq \lambda } \frac { |{\lambda ^\prime }\rangle \langle {\lambda ^\prime }| 
\hat {\cal W} | {\lambda }\rangle \langle {\lambda }| }{E_\lambda ^{(0)} -E_{\lambda ^\prime } ^{(0)} }.
\label{w.17} 
\end{equation} 

Instead of transforming wave functions, we can transform operators: 
\begin{eqnarray} 
\hat{{\tilde{\cal A}}}&=&e^{-\hat{R}}\hat{\cal A} e^{\hat{R}} \nonumber \\ 
&=& \hat{\cal A}-[\hat{ R},\hat{\cal A}]+\frac 12[\hat{ R},[\hat{ R},\hat{\cal A}]]+\dots \; . 
\label{w.18} 
\end{eqnarray} 
Here ${\cal A}$ stands for an arbitrary operator. If, for example, we substitute $\hat{\cal H}$ instead of ${\cal A}$ into 
Eq.~(\ref{w.18}) and apply Eq.~(\ref{w.17}), we obtain 
\begin{equation} 
\hat{\tilde{\cal H }}=\hat{\cal H}_0-\varepsilon ^2 \Delta \hat{\cal H}_2 + \dots \; , 
\label{w.19} 
\end{equation} 
where $\Delta \hat{\cal H}_2= [\hat{R}_2,\hat{\cal H}_0]+[\hat{R}_1,\hat{\cal W}]-
\frac 12 [\hat{ R}_1,[\hat{ R}_1,\hat{\cal H}_0]] $ is the term describing second-order correction to the 
unperturbed energies and, in a case of coupling to continuum,  widths (to the second-order approximation) 
of decaying states. The term 
linear in $\varepsilon $ is absent because $\langle \lambda |\hat{\cal W}|\lambda \rangle =0$ for 
all $\lambda $ by assumption. However, the significance of Eq. (\ref{w.18}) transcends far beyond derivation of 
eigenenergies $E_\lambda $. Eq. (\ref{w.18}) provides the transformation rule for any arbitrary operator, 
in particular, the atomic 1D density operator. The first-order approximation for Eq. (\ref{w.18}) reads as 
\begin{equation} 
\hat{{\tilde{\cal A}}}\approx \hat{\cal A}-\varepsilon [\hat{ R}_1,\hat{\cal A}] , 
\label{w.18bis} 
\end{equation} 
where $\hat{R}_1$ is given by Eq. (\ref{w.17}). 

\subsection{Application to the Lieb-Liniger model perturbed by three-body collisions in the weakly-interacting regime}
\label{subsec:2.2} 
We take now $\hat{\cal H}_0=\hat{H}_\mathrm{h}$ and $\varepsilon \hat{\cal W}=\hat{H}_\mathrm{a}$, where the anharmonic part of 
the Hamiltonian is approximated by Eq. (\ref{w.12}). Direct calculation  of the corresponding $\hat{R}\approx \varepsilon \hat{R}_1$ is 
rather involved, therefore we use the following approach. First, we note that $\hat{R}_1$ is cubic in creation (annihilation) operators 
of elementary excitations, being the eigenmodes of $\hat{H}_\mathrm{h}$. This means that the general form of $\varepsilon \hat{R}_1$ is 
also cubic in terms of ``bare" phase and density operators: 
\begin{eqnarray} 
\varepsilon \hat{R}_1\equiv &\hat{\cal R}_1&= \frac i{\sqrt{L}} \sum _{q,q^\prime } \big{(} B^{-q-q^\prime \; q \;q^\prime } 
\hat{\phi }_{q } \hat{\phi }_{-q-q^\prime } \hat{\phi }_{q^\prime } +  \nonumber \\ && 
B_{-q-q^\prime }^{q \;q^\prime } 
\hat{\phi }_{q } \hat{n }_{-q-q^\prime } \hat{\phi }_{q^\prime }+ B^{-q-q^\prime } _{ q \;q^\prime } 
\hat{n}_{q } \hat{\phi }_{-q-q^\prime } \hat{n}_{q^\prime }+ \nonumber \\ && 
B_{-q-q^\prime \; q \;q^\prime } 
\hat{n}_{q } \hat{n}_{-q-q^\prime } \hat{n}_{q^\prime } \big{)} . 
\label{w.20} 
\end{eqnarray} 
The summation in  Eq. (\ref{w.20}) is taken over non-zero wavenumbers, 
i.e., if one of the wavenumbers $q$, $q^\prime $ or $q+q^\prime $ equals to 0, then the 
corresponding term is dropped off the sum. 

It is easy to prove that the general solution of an operator equation 
$[ \hat{\cal Q}, \hat{\cal H}_0]=\varepsilon \hat{\cal W}$, which follows from 
Eq. (\ref{w.19}), is $\hat{\cal Q}=\varepsilon \hat{R}_1 +\hat{\cal I}$, 
where $\hat{\cal I}$ is any operator commuting with the unperturbed Hamiltonian, $[\hat{\cal I},\hat{\cal H}_0]$, i.e., 
corresponding to a certain integral of motion. However, because of energy and momentum conservation in 1D, there is no operator, 
which has the form of Eq. (\ref{w.20}) and commutes with $\hat{H}_\mathrm{h}$. In other words, the 
sought for polaronic unitary transformation can be uniquely determined (to the linear order) 
from the requirement  of the perturbation operator being cancelled by this transformation. Therefore 
we transform the density operator according to 
\begin{equation} 
\hat{\tilde{n}}_k=\hat{n}_k-[\hat{\cal R}_1, \hat{{n}}_k], 
\label{w.23}
\end{equation} 
where $\hat{\cal R}_1$ is the  solution of the operator equation 
\begin{equation} 
[\hat{\cal R}_1, \hat{H}_\mathrm{h} ]=\hat{H} _\mathrm{a} 
\label{w.24} 
\end{equation}
under the constraint (\ref{w.20}), that provides correct structure of the correct solution and ensures its uniqueness. 

Now solve Eq. (\ref{w.24}), recalling the explicit form of the unperturbed Hamiltonian Eq	(\ref{w.8}) and the perturbation 
operator Eq. (\ref{w.12}). We note  that, obviously, $B_{-q-q^\prime }^{q \;q^\prime }\equiv 0$ and 
$B_{-q-q^\prime \; q \;q^\prime }\equiv 0$. Then symmetry arguments allow us to state that the coefficient 
$B^{{-q-q^\prime }\; q \;q^\prime } $ does not change under any permutation of its arguments and 
$B^{{-q-q^\prime }} _{ q \;q^\prime }=B^{{-q-q^\prime }} _{ q^\prime \;  q}$. After some algebra we obtain a set of 
equations for these non-zero coefficients, which in the matrix form reads as 
\begin{widetext} 
\begin{equation} 
\left( 
\begin{array}{cccc} 
-3 [  (k-q)^2/4 +1] &      q^2        &          k^2    &     0           \\ 
-3 (  q^2/4+1)      &  (k-q)^2        &          0      &    k^2          \\ 
-3 (  k^2/4+1)      &       0         &       (k-q)^2   &    q^2          \\
0                   & -(k^2/4+1)      & -(q^2/4+1)      & -[(k-q)^2/4+1]
\end{array}
\right) 
\left( 
\begin{array}{l} 
n_\mathrm{1D} ^{-1} B^{k-q\; -k\; q}  \\
n_\mathrm{1D}       B^{-k} _{k-q\; q} \\
n_\mathrm{1D}       B^{q} _{k-q\; -k} \\
n_\mathrm{1D}       B^{k-q} _{-k\; q} 
\end{array} 
\right) = 
\left( 
\begin{array}{c} 
qk/2           \\
k(k-q)/2       \\
-q(k-q)/2      \\ 
-(k^2-kq+q^2)/8
\end{array} 
\right) . 
\label{w.25} 
\end{equation}
\end{widetext} 
The solution of the set of Eqs. (\ref{w.25}) is 
\begin{eqnarray} 
B^{k-q\; -k\; q}  &=& \frac {n_\mathrm{1D}}3, \label{w.22} \\
B^{-k} _{k-q\; q} &=&\frac 1{n_\mathrm{1D}} \left[ \frac 14 -\frac 1{q(k-q)} \right] , \label{w.22bis} \\
B^{q} _{k-q\; -k} &=&\frac 1{n_\mathrm{1D}} \left[ \frac 14 +\frac 1{k(k-q)} \right] , \label{w.26}    \\
B^{k-q} _{-k\; q} &=&\frac 1{n_\mathrm{1D}} \left( \frac 14 +\frac 1{kq} \right) . \label{w.26bis} 
\end{eqnarray} 

Then we find the explicit form of the transformation Eq. (\ref{w.23}) 
\begin{eqnarray} 
\hat{\tilde{n}}_k&=&\hat{n}_k+\frac 1{\sqrt{L}} \sum _q \bigg{ \{ }n_\mathrm{1D} \hat{\phi } _{k-q}\hat{\phi } _{q}+ \nonumber \\ &&
\frac 1{n_\mathrm{1D}} \left[ \frac 14 -\frac 1{q(k-q)} \right] \hat{n} _{k-q}\hat{n} _{q} \bigg{ \} } . 
\label{w.27} 
\end{eqnarray} 

\section{Transition matrix element}
\label{sec:3}
We transform the integrability-breaking interaction term Eq. (\ref{w.11}) by changing $\hat{n}_k$ to $\hat{\tilde{n}}_k$, applying 
Eq.~(\ref{w.27}), and retaining, according to the assumed order of approximation, only quartic terms. Then after some algebra we 
obtain the matrix element $\langle k_0-k_1-k_2,\, k_1,\, k_2 | \hat{\tilde{H}} _\mathrm  {ni}|k_0\rangle $, where $|k_0\rangle = 
\hat{b}_{k_0}^\dag |0\rangle $,  $|k_0-k_1-k_2,\, k_1,\, k_2\rangle = 
\hat{b}_{k_0-k_1-k_2}^\dag \hat{b}_{k_1}^\dag \hat{b}_{k_2}^\dag |0\rangle $, and $|0\rangle $ is the vacuum of 
Bogoliubov elementary excitations. In a case of arbitrary initial numbers of the involved elementary 
excitations, the relevant matrix element 
\begin{eqnarray} 
\langle \mathrm{f}|   \hat{\tilde{H}} _\mathrm  {ni}|\mathrm{in}\rangle  &= &  
\sqrt{(n_{k_1} +1)(n_{k_2} +1)(n_{k_0-k_1-k_2} +1)n_{k_0}} \times  \nonumber \\ 
&&\langle k_0-k_1-k_2,\, k_1,\, k_2 | \hat{\tilde{H}} _\mathrm  {ni}|k_0\rangle , \label{w.38} 
\end{eqnarray} 
where 
\begin{eqnarray} 
|\mathrm{in} \rangle &=& (n_{k_1} !n_{k_2} !n_{k_0-k_1-k_2}!n_{k_0}!)^{-1/2} \times \nonumber \\ && 
\hat{b}_{k_0}^{\dag \, n_{k_0}} 
\hat{b}_{k_0-k_1-k_2}^{\dag \, n_{k_0-k_1-k_2}}\hat{b}_{k_1}^{\dag \, n_{k_1}}\hat{b}_{k_2}^{\dag \,n_{k_2}}|0\rangle , 
\label{w.39} \\ 
|\mathrm{f} \rangle &=& [(n_{k_1}+1)!(n_{k_2}+1)!\times \nonumber \\ && 
(n_{k_0-k_1-k_2}+1)!(n_{k_0}-1)!]^{-1/2} \times \nonumber \\ && 
\hat{b}_{k_0}^{\dag \, n_{k_0}-1} 
\hat{b}_{k_0-k_1-k_2}^{\dag \, n_{k_0-k_1-k_2}+1}\hat{b}_{k_1}^{\dag \, n_{k_1}+1}\hat{b}_{k_2}^{\dag \,n_{k_2}+1}|0\rangle ,  
\label{w.39bis} 
\end{eqnarray} 
is readily expressed through $\langle k_0-k_1-k_2,\, k_1,\, k_2 | \hat{\tilde{H}} _\mathrm  {ni}|k_0\rangle $ and matrix 
elements of the bosonic field operators. Eq. (\ref{w.38}) enables us to evaluate the transition rates in a quasicondensate at 
non-zero temperature. 

After some algebra we obtain 
\begin{eqnarray}
\langle k_0-k_1-k_2,\, k_1,\, k_2 | \hat{\tilde{H}} _\mathrm  {ni}|k_0\rangle = \qquad && \nonumber \\
\qquad -\frac {12\kappa _\mathrm{3b}}{Ln_\mathrm{1D}} 
{\cal M}^{ k_0-k_1-k_2 \, k_1 \, k_2 }_{k_0}, && 
\label{w.28} 
\end{eqnarray} 
where the dimensionless form of the matrix element is 
\begin{widetext} 
\begin{eqnarray} 
{\cal M}^{ k_0-k_1-k_2\, k_1\, k_2 }_{k_0}&=&\sqrt{S_{k_0-k_1-k_2}S_{k_2}}\, {\cal Y}_{k_0\, k_1} + 
\sqrt{S_{k_0-k_1-k_2}S_{k_1}}\, {\cal Y}_{k_0\, k_2}+\sqrt{S_{k_1}S_{k_2}}\, {\cal Y}_{k_0\, k_0-k_1-k_2}+ \nonumber \\ && 
\sqrt{S_{k_0}S_{k_1}}\, {\cal Z}_{k_0-k_1-k_2\, k_2}+ \sqrt{S_{k_0}S_{k_2}}\, {\cal Z}_{k_0-k_1-k_2\, k_1}+ 
\sqrt{S_{k_0}S_{k_0-k_1-k_2}}\, {\cal Z}_{k_1\, k_2} ,   
\label{w.28bis} 
\end{eqnarray} 
\end{widetext}
\begin{eqnarray}
{\cal Y}_{q\, q^\prime }&=&\frac 1{4\sqrt{S_qS_{q^\prime }}}+\left( \frac 14 -\frac 1{qq^\prime }\right) \sqrt{S_qS_{q^\prime }} , 
\label{w.29} \\ 
{\cal Z}_{q\, q^\prime }&=&-\frac 1{4\sqrt{S_qS_{q^\prime }}}+\left( \frac 14 -\frac 1{qq^\prime }\right) \sqrt{S_qS_{q^\prime }} . 
\label{w.29bis}
\end{eqnarray} 

The significance of our method to derive Eqs. (\ref{w.28}~-- \ref{w.29bis}) can be understood at the best from comparison to a 
``naive" alternative way of derivation. Namely, we may  express the density operator in Eq. (\ref{w.5}) through the bosonic 
field creation and annihilation operators as $\hat{n} =\hat{\psi }^\dag \hat{\psi }$. Although there is no true condensate in 
the thermodynamic limit in 1D, we 
may try, by integrating out slow variables \cite{Popov} (up to the infrared cut-off momentum $k_\mathrm{IR}$), 
to express the bosonic field operator as 
$\hat{\psi } =\sqrt{n_\mathrm{1D}} +\delta \psi $, where $\delta \psi =L^{-1/2} \sum _{|k|>k_\mathrm{IR}} \hat{a}_k e^{ikz} $ 
and the bare atomic operators $\hat{a}_k$ 
in the momentum representation are related to the quasiparticle operators $\hat{b}_k$, $\hat{b}_{-k}^\dag $ via the Bogoliubov 
transformation. The correct value of the 
rate of Beliaev and Landau damping in weakly interacting two-dimensional Bose gases (where true condensate is absent at any finite 
temperature) can be found in this way \cite{BA}, therefore it is not clear from 
the very beginning that this method fails for three-body collisions in 1D. After renormalizing the two-body interaction strength 
and singling out the integrability-breaking terms terms one could obtain, to the lowest order in $n_\mathrm{1D} a_s$ an 
incorrect form for the integrability-breaking interactions 
\begin{equation} 
\hat{H}_\mathrm{ni}= -\frac {3\kappa _{3b}}{n_\mathrm{1D}} : \int _0 ^L dz\, (\delta \psi ^\dag +\delta \psi )^2
\delta \psi ^\dag \delta \psi : 
\label{w.5bis} 
\end{equation} 
(here symbol :~: denotes normal ordering of the atomic field operators), 
that yields the result similar 
to Eqs. (\ref{w.28},~\ref{w.28bis}), but with ${\cal Y}_{q\, q^\prime }$, ${\cal Z}_{q\, q^\prime }$ substituted by 
\begin{eqnarray}
{\cal Y}_{q\, q^\prime }^\mathrm{(sg)}&=&\frac 1{4\sqrt{S_qS_{q^\prime }}}+\frac { \sqrt{S_qS_{q^\prime }}}4 , 
\label{w.30} \\ 
{\cal Z}_{q\, q^\prime }^\mathrm{(sg)}&=&-\frac 1{4\sqrt{S_qS_{q^\prime }}}+  \frac {\sqrt{S_qS_{q^\prime }}}4 , 
\label{w.30bis}
\end{eqnarray}
respectively. As we see later, Eqs. (\ref{w.30}, \ref{w.30bis}) lead to a singularity of the matrix element 
${\cal M}^{ k_0-k_1-k_2\, k_1\, k_2 }_{k_0}$ at vanishing momentum transfer, thus signifying the 
inapplicability of the ``naive" approach.

To the end of this Section, we discuss the kinematics of the three-body process shown in Fig.~\ref{F1} and the 
values of the matrix element (\ref{w.28bis}) on the energy shell, where the energy conservation law requires 
\begin{equation}
\epsilon _{k_0} = \epsilon _{k_1}+\epsilon _{k_2}+\epsilon _{k_0-k_1-k_2}            \label{w.31} 
\end{equation} 
and the energy of an elementary excitation is given by Eq. (\ref{w.11bis}). Two momenta of the 
product elementary excitations (we denote them 
by $k_1$ and $k_0-k_1-k_2$, assuming for convenience $|k_1|<|k_0-k_1-k_2|$) have the same sign as $k_0$, and 
the third excitation momentum has the opposite sign, $\mathrm{sgn} \, (k_2 k_0) <0$. To be definite, we assume $k_0>0$. 

\subsection{Phononic limit} 
\label{sec:3.1} 
If $k_0\lesssim 1$ then all relevant excitations are phonons having almost linear dispersion law. Only taking into account 
the cubic correction $\epsilon _k\approx |k| +|k|^3/8$ helps us to find from Eq. (\ref{w.31}) the relation between 
the momenta of the phonons:  
\begin{equation} 
k_2\approx -\frac 3{16} k_0(k_0-k_1)k_1   .  \label{w.32} 
\end{equation}      

The matrix element (\ref{w.28}) calculated using Eqs. (\ref{w.29}, \ref{w.29bis}) in the phononic limit is 
\begin{equation} 
{\cal M}^{ k_0-k_1-k_2\, k_1\, k_2 }_{k_0}\approx \frac 12 \sqrt{\frac {k_1(k_0-k_1-k_2)}{k_0|k_2|}} . 
\label{w.33} 
\end{equation} 
On the energy shell, where Eq. (\ref{w.32}) holds, Eq. (\ref{w.33}) is reduced to 
\begin{equation} 
{\cal M}^{ k_0-k_1-k_2\, k_1\, k_2 }_{k_0}\approx \frac 2{\sqrt{3}k_0} . 
\label{w.33bis} 
\end{equation}
The matrix element (\ref{w.33bis}) is finite in the limit of vanishing transferred momentum, $k_1\rightarrow 0$, 
unlike the incorrect ``naive" estimation that relies on Eqs. (\ref{w.30}, \ref{w.30bis}) and predicts the matrix element 
diverging as $1/k_1$. 

However, this process alone is suppressed for low-energy excitations (phonons) in trapped condensates. 
The reason is the infrared momentum cut-off imposed by a finite length $\ell $ of a trapped 1D quasicondensate. 
From the condition $|k|_2\gtrsim 2\pi /\ell $ and Eq. (\ref{w.32}) we conclude that this process lead to damping 
of phonons with $k_0\gtrsim 5\, \ell ^{-1/3}$ (or, in usual units, $k_0\gtrsim  5\, \zeta _\mathrm{h}^{-2/3}\ell ^{-1/3}$. 
In a $^{87}$Rb quasicondensate 
with the density $n_\mathrm{1D}=50~\mu \mathrm{m}^{-1}$, length $\ell =100~\mu $m, and 
the radial trapping frequency $\omega _\perp = 2\pi \times 3$~kHz the process shown in Fig.~\ref{F1} is kinematically 
allowed for $k_0\gtrsim 3.5\times 10^4~\mathrm{cm}^{-1}$, which is very close to the crossover $k_0 \sim 1/\zeta _\mathrm{h}$
between the phononic and particle-like parts of the Bogoliubov excitation spectrum.  

\begin{figure} 
\vspace*{1mm}
\begin{center}
\epsfig{file=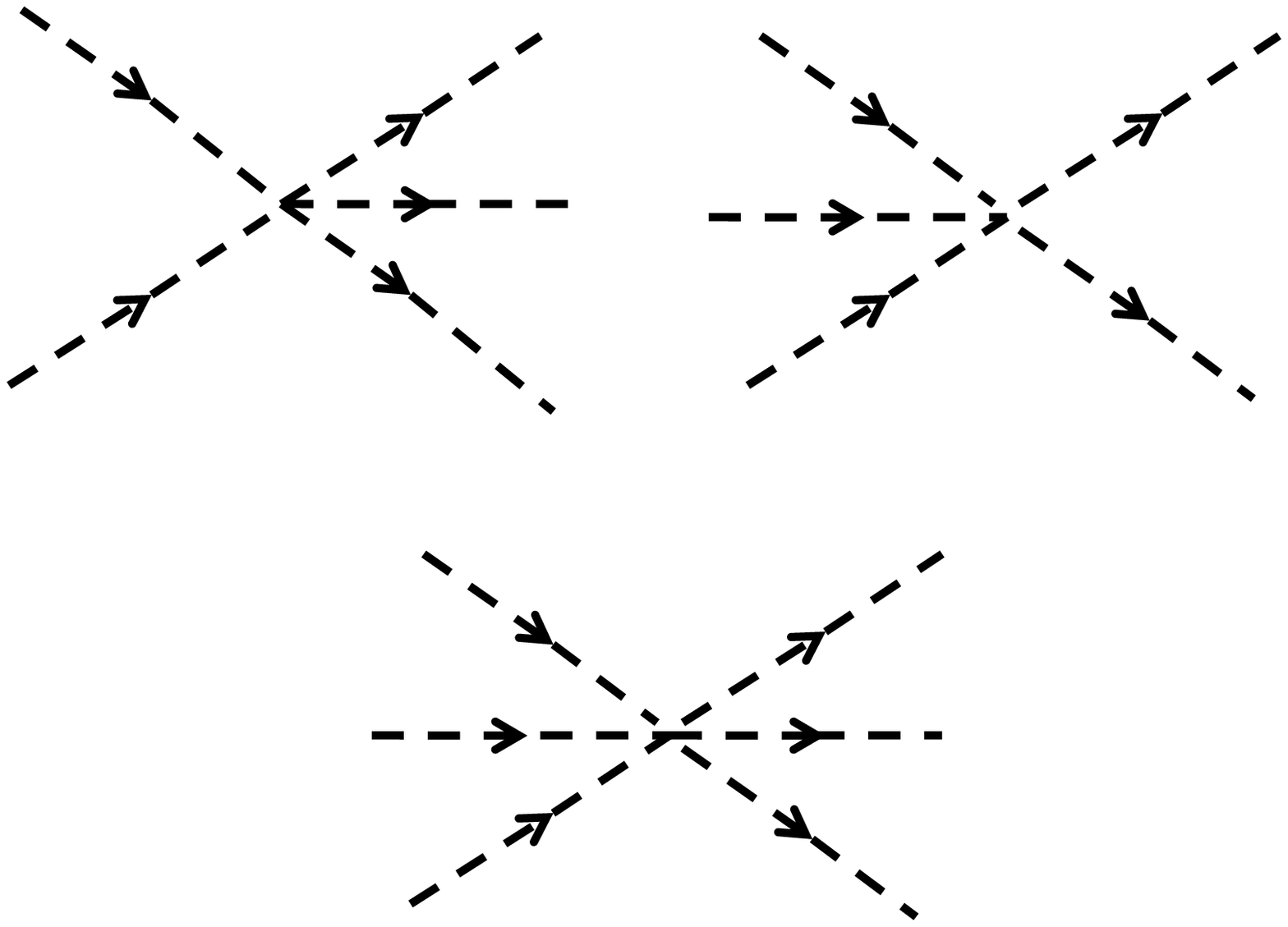,width=65mm} 
\end{center} 
\begin{caption}
{Feynman diagrams of the processes providing thermalization of low-energy excitations (phonons). 
\label{F2} }
\end{caption}
\end{figure}

The processes shown in Fig.~\ref{F2}, in contrast, have less severe kinematic restriction, $k_0\gtrsim 2\pi /\ell $, and 
therefore can thermalize elementary excitations deeply in the phononic regime. 

\begin{figure}

\vspace*{5mm} 

\epsfig{file=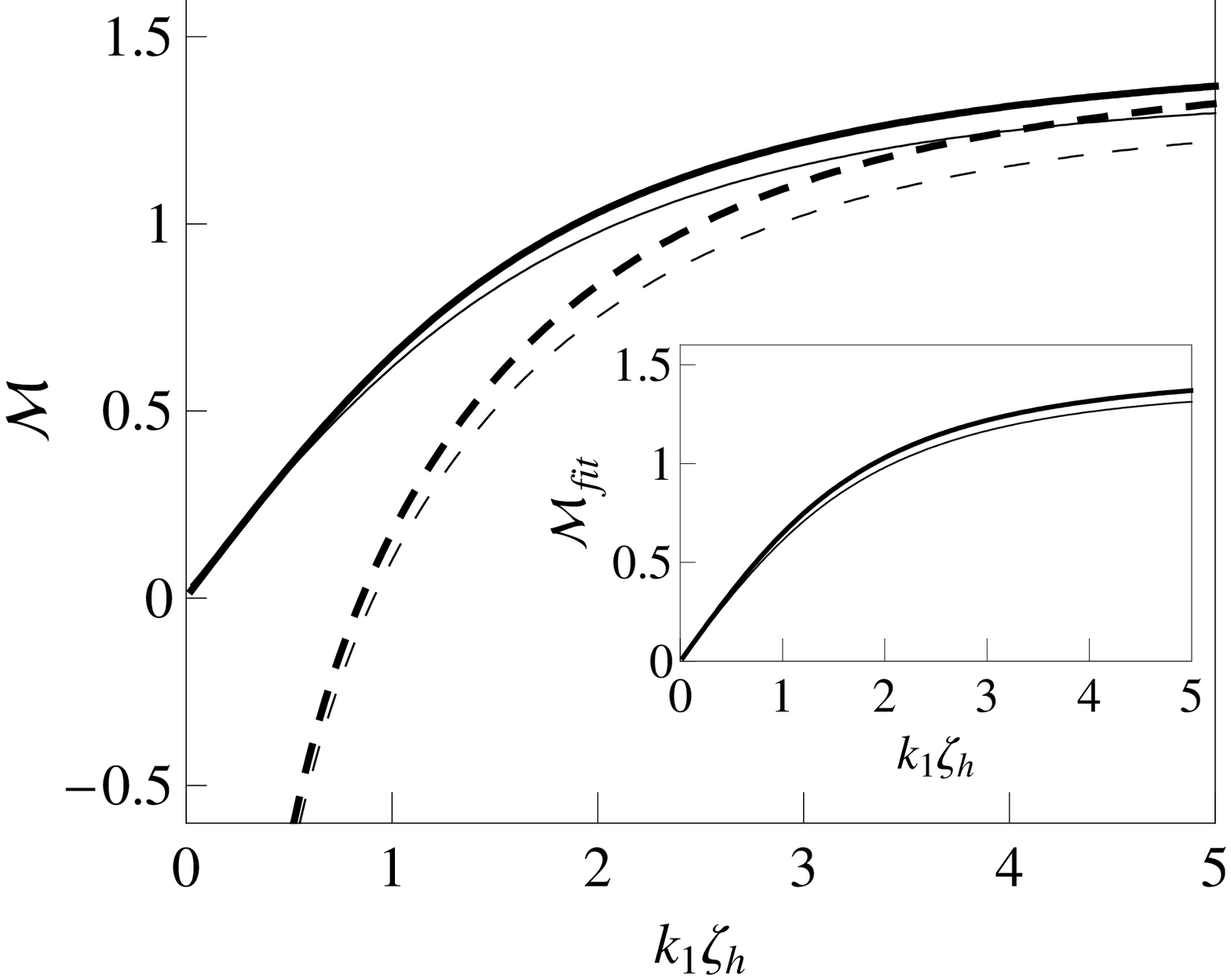,width=\columnwidth } 

\begin{caption} 
{Transition matrix element ${\cal M}\equiv {\cal M}^{ k_0-k_1-k_2\, k_1\, k_2 }_{k_0}$ as a function of the 
product state wavenumber $k_1$. Units on axes are dimensionless. On the horizontal axis label we explicitly 
indicate scaling of $k_1$ to the inverse healing length. ${\cal M}$ is calculated on the energy shell, i.e. the energy conservation 
Eq. (\ref{w.31}) holds. Solid lines: the main result of Sec. \ref{sec:3}, given by Eqs. (\ref{w.28bis}~--~\ref{w.29bis}). 
Dashed lines: approximation by Eqs. (\ref{w.30},~\ref{w.30bis}) (singular at $k_1\rightarrow 0$). Initial momentum 
$k_1=10\, \zeta _\mathrm{h} ^{-1}$ (thin lines) and $25\, \zeta _\mathrm{h} ^{-1}$ (thick lines). Fitting by Eq. (\ref{w.33bis}) is 
practically indistinguishable from the solid lines and is  displayed in the inset.
\label{F3} }
\end{caption}  

\end{figure} 

\subsection{Fast-particle limit} 
\label{sec:3.2} 
In this limit $k_0\gg 1$. As concerns the the product states, there are two distinct cases \cite{Glazman}. 
All three product elementary excitations can be fast as well, in this case Eq.(\ref{w.31}) yields 
\begin{equation}   
k_2 = \frac {k_0-k_1}2 - \frac 12\sqrt{k_0^2+2k_0k_1-3k_1^2} . 
\label{w.34}
\end{equation} 
The aforementioned ordering convention $k_2<0$, $0<k_1<k_0-k_1-k_2$, $k_0>0$ now results in 
\begin{equation} 
0<k_1<2k_0/3.             \label{w.35} 
\end{equation} 

But there is another possibility, when two product elementary excitations are phonons, $k_1\lesssim 1$ and 
\begin{equation} 
k_2=-\frac {k_0-1}{k_0+1}k_1 .           \label{w.36} 
\end{equation} 
Although the phase space corresponding to the latter case is relatively small, collisions with small 
transferred momentum are bosonically amplified in a quantum gas with high thermal population of phononic 
modes, and their contribution to the fast particle damping can be significant \cite{Glazman}. 

Eqs. (\ref{w.28bis} -- \ref{w.29bis}) yield on the energy shell, i.e. for Eqs. (\ref{w.34}) and (\ref{w.36}) 
holding for $k_1\gg 1$ and $k_1\lesssim 1$, respectively
\begin{equation} 
{\cal M}^{ k_0-k_1-k_2\, k_1\, k_2 }_{k_0}=\left \{ 
\begin{array}{ll} 
3 \sqrt{|k_1 k_2|}/4 \, , & k_1\lesssim 1 \\ 
3/2 \,  ,& k_1 \gg 1 
\end{array} 
\right. \; . 
\label{w.37}     
\end{equation} 
The formula interpolating between these two cases is 
\begin{equation} 
{\cal M}^{ k_0-k_1-k_2\, k_1\, k_2 }_{k_0}=\frac 32 \sqrt{S_{k_1} S_{k_2} }, 
\label{w.37bis}     
\end{equation} 
in full agreement with the  used in Ref. \cite{Glazman} assumption of correlation 
separation 
\begin{equation} 
\langle k_0-k_1-k_2\, k_1\, k_2 |\hat{n}^3|k_0\rangle = 3\langle  k_1\, k_2 |\hat{n}^2|0\rangle 
\langle k_0-k_1-k_2|\hat{n}| k_0\rangle 
\label{w.40}     
\end{equation} 
that singles out the fast particle contribution.

In Fig. \ref{F3} we display the matrix element Eq. (\ref{w.28bis}) calculated in the case of $k_0\gg 1$ 
with ${\cal Y}_{q\, q^\prime }$, ${\cal Z}_{q\, q^\prime }$ given by Eqs. (\ref{w.29},~\ref{w.29bis}), and, 
for comparison, its value calculated from Eqs. (\ref{w.5bis}~--~\ref{w.30bis}). We see that the latter 
approach works only for asymptotically large $k_1$, and the former one is in a very good agreement with 
the approximation (\ref{w.37bis}). 

Approximation (\ref{w.37bis}) breaks down out of the energy shell. This can be important for transient processes, with the 
energy uncertainty inversely proportional to the duration of the external driving of the system, like in the experiment \cite{NirD1}. 
However, this problem (also related to the quantum Zeno and anti-Zeno effects \cite{KKNat}) 
is to be considered separately from the kinetic equation, which follows from Fermi's golden rule and is 
considered in the next Section. 

\section{Kinetic equation} 
\label{sec:4} 
\subsection{General approach} 
\label{sec:4.0} 
Using Fermi's golden rule, we can readily write a 1D kinetic equation that takes into account effects of bosonic amplification 
[cf. Eq. (\ref{w.38})]:
\begin{widetext} 
\begin{eqnarray} 
\frac \partial {\partial t}f_k &=& -\gamma _0 \bigg \{ \int \int _{Q_{12}} dk_1dk_2 \, \delta (
\epsilon _{k-k_1-k_2}+\epsilon _{k_1}+\epsilon _{k_2} -\epsilon _{k} ) 
\left| {\cal M}^{ k-k_1-k_2\, k_1\, k_2 }_{k}\right| ^2 \times                        \nonumber \\ && 
[(f_{k-k_1-k_2}+1)(f_{k_1}+1)(f_{k_2}+1)f_{k}-f_{k-k_1-k_2}f_{k_1}f_{k_2}(f_{k}+1)] + \nonumber \\ && 
\int \int _{Q_{12}^\prime } dk_1dk_2 \, \delta (
\epsilon _{k}+\epsilon _{k_1}+\epsilon _{k_2} -\epsilon _{k+k_1+k_2} ) 
\left| {\cal M}^{ k\, k_1\, k_2 }_{k+k_1+k_2}\right| ^2 \times                        \nonumber \\ && 
[ f_{k}f_{k_1}f_{k_2}(f_{k+k_1+k_2}+1)-(f_{k}+1)(f_{k_1}+1)(f_{k_2}+1)f_{k+k_1+k_2} ]\bigg \} . 
\label{w.41} 
\end{eqnarray} 
\end{widetext} 
Here $f_k$ is the population of the elementary excitation mode with the 1D momentum $k$. 
The rate of the process is scaled to 
$\gamma _0 =(72/\pi )( \kappa _{\mathrm{3b} }/n_{\mathrm{1D} })^2$ or, in usual units, 
\begin{equation} 
\gamma _0= \frac {18}\pi \omega _\perp \xi ^2 \left( \frac {n_{\mathrm{1D}}a_s^2}{l_\perp } \right) ^2 , 
\label{w.42} 
\end{equation} 
where $l_\perp =\sqrt{\hbar /(m\omega _\perp )}$ is the radial trapping length scale. The first term in curly brackets 
in Eq. (\ref{w.41}) corresponds to processes shown in Fig.~\ref{F1} with the wave number $k$ assigned to the (single) 
incoming line in 
Fig.~\ref{F1}(a) or the (single) outgoing line in Fig.~\ref{F1}(b). The second term describes the situation of $k$ being 
one of the three incoming quasiparticle momenta in  Fig.~\ref{F1}(b)  or one of the three outgoing quasiparticles in 
Fig.~\ref{F1}(a). The respective integration ranges take into account the bosonic nature of  elementary excitations and are 
defined in a way that prevents double counting of the same momenta of elementary excitations. 
$Q_{12}$ includes all $k_1$, $k_2$ satisfying 
the conditions $ \mathrm{sgn} \, k_1 = \mathrm{sgn} \,(k-k_1-k_2)=
- \mathrm{sgn} \, k_2= \mathrm{sgn} \, k$, $|k_1|<|k-k_1-k_2|$, $0<|k_2|<\infty $; $Q_{12}^\prime $ consists of two regions: 
(i) $ \mathrm{sgn} \, k = \mathrm{sgn} \, k_1 =- \mathrm{sgn} \,k_2 = \mathrm{sgn} \,(k-k_1-k_2)$, 
$0<|k_1|<\infty $, $0<|k_2|<\infty $, 
and $ \mathrm{sgn} \, k =- \mathrm{sgn} \, k_1 =- \mathrm{sgn} \,k_2 =- \mathrm{sgn} \,(k-k_1-k_2)$, 
$0<|k_1|<|k-k_1-k_2|<\infty $.

In trapped quasicondensates, Eq. (\ref{w.41}) applies in the local density approximation if discreteness of the 
elementary exciation spectrum is not important, i.e., if $|k_2|\gg 2\pi /\ell $. Practically, this means Eq. (\ref{w.41}) 
applies in the fast particle damping regime for scattering events with $|k_1|\approx |k_2|\gg 2\pi /\ell $. 
In this case, competing processes shown in Fig.~\ref{F2} should provide relatively small contribution to the thermalization rate, 
however, their detailed calculation is to be a subject of future work. A rough estimation (based on evaluation of 
phase space available for scattering of a fast particle with wave number $k\gg 1$ on other quasiparticles)  
predicts that the diagrams from the 
first and second raws yield the rates smaller by a factor $k/n_\mathrm{1D}$ and $(k/n_\mathrm{1D})^2$, respectively. 

It is easy to check that the stationary solution of Eq. (\ref{w.41}) is the Bose-Einstein equilibrium distribution 
\begin{equation} 
f^\mathrm{e} _k=\frac 1{\exp (\epsilon _k /T)-1} .
\label{w.43} 
\end{equation} 
Here we measure temperature $T$ in energy units (i.e., set Boltzmann's constant to 1). Eq. (\ref{w.41}) does not conserve the total 
number of elementary excitations, therefore chemical potential in the Bose-Einstein distribution   (\ref{w.42}) is zero. 
 
If the energy $\epsilon _k$ of the fast particle is much larger than both the temperature and mean interaction energy per particle 
then  scattering  with large transferred momentum populates initially empty particle-like modes, and the population of the 
$k$-mode decreases exponentially with the decrement $\Gamma ^\mathrm{damp} =\sqrt{3} \pi \gamma _0/4$ \cite{Mazets2,Glazman} 
(note that $\Gamma ^\mathrm{damp}$ does not depend on $k$).  

It has been noticed \cite{Glazman} that three-body collisions with small momentum transfer can give, due to  bosonic amplification, 
major contribution to the damping rate of a fast particle in a certain parameter range. In the present paper we make a step 
forward compared to the  treatment of Ref. \cite{Glazman} by looking precisely to the dynamics of a bunch of fast particles in a 
1D quasicondensate. 

\subsection{Fokker-Planck equation} 
\label{sec:4.1} 
We consider a non-equilibrium distribution of elementary excitations $f_k=f_k^\mathrm{e}+\delta f_k$, where the perturbed part is 
small, $\int _{-\infty }^\infty dk\, \delta f_k\ll \int _{-\infty }^\infty dk\, f_k^\mathrm{e}$, and is localized in the 
momentum space around $k_0\gg 1$ (the width of the fast particles bunch being $\Delta k\ll k_0$). Such a distribution can be 
created by means of Bragg spectroscopy \cite{Bragg1}, perhaps, using higher-order processes \cite{Bragg2} to obtain large 
values of $k_0$. To be still in the 1D regime and avoid scattering of atoms to radially excited states we must require the
kinetic equation of the fast particles to be lees than $2\hbar \omega _\perp $ (factor 2 appears here due to parity conservation 
\cite{Mazets1,Mazets2}). A relatively small numbers of fast particles allows us to neglect heating of the lower modes in the 
course evolution.

In what follows we take into account only collisions with small momentum transfer. Then we analyze the dynamics of $\delta f_k$ on 
the time scale much shorter than $1/\Gamma ^\mathrm{damp} $ (when we can neglect scattering with large momentum transfer). 
A change of $\delta f_k$  on such a short time scal  will be a measure of the importance of the  scattering  
events with the small momentum transfer, bosonically-amplified 
by $f_{k_1}^\mathrm{e}\sim f_{k_2}^\mathrm{e}\gg 1$.

The assumptions listed above enable us to linearize the kinetic equation (\ref{w.41}) and  reduce it to the Fokker-Planck equation 
\cite{Gardiner}. We neglect the momentum dependence of the kinetic coefficients (diffusion $D$ and advection $A$ within the 
narrow momentum around $k_0$ and finally obtain 
\begin{equation} 
\frac \partial {\partial t} \delta f_k =A \frac \partial {\partial k}\delta f_k+D\frac {\partial ^2}{\partial k^2}\delta f_k , 
\label{w.44} 
\end{equation} 
\begin{widetext} 
\begin{eqnarray} 
A&=& \gamma _0 \int _0^\infty dk_1 \int _{-\infty}^0 dk_2\, (k_1+k_2)
\left| {\cal M}^{ k_0-k_1-k_2\, k_1\, k_2 }_{k_0}\right| ^2 \delta (
\epsilon _{k_0-k_1-k_2}+\epsilon _{k_1}+\epsilon _{k_2} -\epsilon _{k_0} )(f_{k_1}^\mathrm{e} +f_{k_2}^\mathrm{e}), 
\label{w.45a} \\ 
D&=& \gamma _0 \int _0^\infty dk_1 \int _{-\infty}^0 dk_2\, (k_1+k_2)^2 
\left| {\cal M}^{ k_0-k_1-k_2\, k_1\, k_2 }_{k_0}\right| ^2 \delta (
\epsilon _{k_0-k_1-k_2}+\epsilon _{k_1}+\epsilon _{k_2} -\epsilon _{k_0} )[f_{k_1}^\mathrm{e}f_{k_2}^\mathrm{e}+
(f_{k_1}^\mathrm{e} +f_{k_2}^\mathrm{e})/2].  ~ 
\label{w.45b}
\end{eqnarray}
\end{widetext}  

Eqs. (\ref{w.44} -- \ref{w.45b}) are written still in dimensionless variables. In what follows we assume usual time and length 
units in Eqs. (\ref{w.44}) and evaluate $A$ and $D$ in two limiting cases. First, for $T\lesssim g_\mathrm{1D}n_\mathrm{1D}$ we 
obtain 
\begin{eqnarray} 
A&\approx &14.6\,  \mathrm{sgn} \, k_0 \, 
\frac {\gamma _0}{|k_0|^2\zeta _{h}^3}\left( \frac {T}{ g_\mathrm{1D}n_\mathrm{1D} }\right) ^4 , 
\label{w.46a} \\ 
D&\approx & 2.45\, \frac {\gamma _0}{|k_0|^3\zeta _{h}^5}\left( \frac {T}{ g_\mathrm{1D}n_\mathrm{1D} }\right) ^5 , 
\label{w.46b}
\end{eqnarray}
where $\gamma _0$ is given by Eq. (\ref{w.42}). 

If, on the contrary, $g_\mathrm{1D}n_\mathrm{1D}\ll T\lesssim 2\hbar \omega _\perp $ then bosonic amplification of 
scattering to free-particle-like mode becomes significant and we obtain 
\begin{eqnarray} 
A&\approx & 14.7\,   \mathrm{sgn} \, k_0 \, 
\frac {\gamma _0}{|k_0|^2\zeta _{h}^3}\left( \frac {T}{ g_\mathrm{1D}n_\mathrm{1D} }\right) ^{3/2} , 
\label{w.47a} \\ 
D&\approx &10.8\, \frac {\gamma _0}{|k_0|^3\zeta _{h}^5}\left( \frac {T}{g_\mathrm{1D}n_\mathrm{1D} }\right) ^{5/2}.  
\label{w.47b}
\end{eqnarray}

The decrease of the kinetic coefficients $A$ and $D$ with growing $k_0$ is explained by two observation. 
First, the total momentum transfer  decreases as $k_0$ increases but the low-energy excitation 
momentum $k_1$ is kept fixed (and $k_2$ is related to $k_0$ and $k_1$ via energy and momentum conservation): 
$k_1+k_2\approx |k_1|(4\zeta _\mathrm{h}^{-2}+k_1^2)^{1/2}/k_0$. Second, integrating in Eqs. 
(\ref{w.45a}, \ref{w.45b}) the delta-function of the energy difference over $k_2$ yields another factor 
$1/k_0$. 

\begin{figure} 

\vspace*{4mm}

\epsfig{file=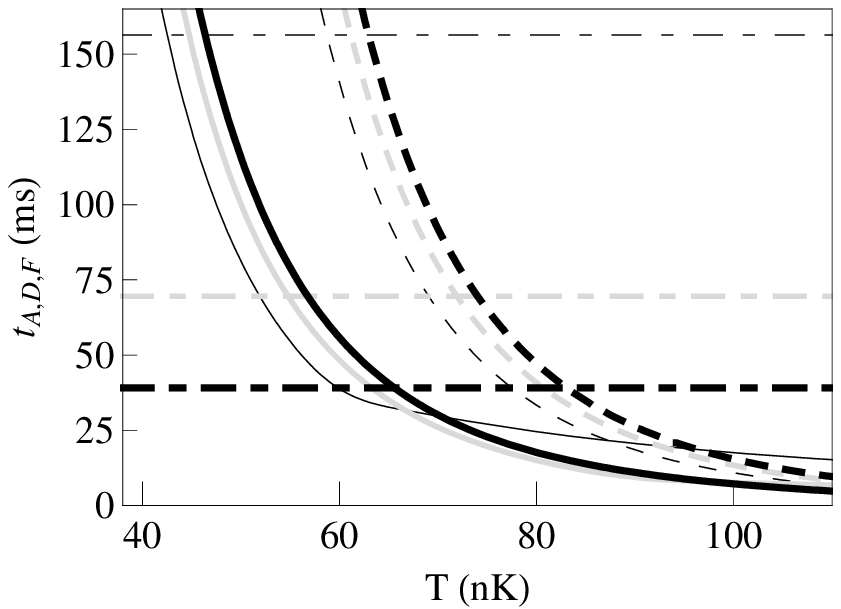,width=\columnwidth } 

\begin{caption} 
{Typical times of shift ($t_A$, solid lines), diffusion ($t_D$, dashed lines), and scattering with large momentum 
transfer ($t_F\equiv 1/\Gamma ^\mathrm{damp}$, dot-dashed lines) for $^{87}$Rb quasicondensate with $n_\mathrm{1D}= 
40\; \mu \mathrm{m}^{-1}$ (thin black lines), $60\; \mu \mathrm{m}^{-1}$ (medium gray lines),   
and $80\; \mu \mathrm{m}^{-1}$ (thick black lines) as functions of temperature. 
$\omega _\perp = 2\pi \times 3$~kHz. Initial distribution of fast atoms is Gaussian with the mean $k_0=\sqrt{2m\omega _\perp /\hbar }=
0.7\times 10^5\; \mathrm{cm}^{-1}$ and the r.m.s. $\Delta k =\frac 14 k_0$. 
\label{F4} }
\end{caption}  

\end{figure}

Assume that the initial distribution of fast particles is Gaussian, centered at $k_0$ and with r.m.s 
(determined, for example, by the duration of the Bragg pulse \cite{Bragg1}) $\Delta k$. The well-known 
solution \cite{Gardiner} of Eq. (\ref{w.44}) with the Gaussian initial condition is $\delta f_k(t)
\propto \exp [ -(k-k_0+At)^2/(2\Delta k^2+4Dt)]/ (2\Delta k^2+4Dt)^{1/2}$. We define the typical time 
scales $t_A$ and $t_D$ as times when shift of the maximum of $\delta f_k$ or increase of its width, 
respectively, become comparable to the initial width $\Delta k$: 
\begin{equation} 
t_A=\Delta k/A, \qquad t_D=\Delta k^2/(2D).                \label{w.48} 
\end{equation} 
Bosonically-amplified three-body collisions with small momentum transfer are practically important, if 
they modify $\delta f_k$ faster, than scattering with large momentum transfer to initially empty modes, i.e., if 
at least one of two inequalities 
\begin{equation} 
\Gamma ^\mathrm{damp} t_A \lesssim 1, \qquad \Gamma ^\mathrm{damp}  t_D \lesssim 1
\label{w.49} 
\end{equation} 
is satisfied. Note, that the number of scattering events per unit time per one fast atom can be 
significantly larger than $\Gamma ^\mathrm{damp} \approx 1.36\, \gamma _0$, but their influence on the 
dynamics of the distribution $\delta f_k$ may be still small.

Fig. \ref{F4} illustrates $t_A$, $t_D$ juxtaposed with $t_F\equiv 1/\gamma ^\mathrm{damp}$ for experimentally realistic system 
parameters. One can see that, at experimentally feasible densities, first the momentum-shift time $t_A$ and then the diffusion time 
$t_D$ become  shorter than the typical time $t_F$, as temperature increases. However, we should always keep in mind 
the limitation $T\lesssim \hbar \omega _\perp $  necessary for the 1D regime. 

Now we can compare our criterion for the significance of the small-transferred-momentum scattering events and that proposed by 
Tan, Pustilnik and Glazman \cite{Glazman}. In the latter case, the small-transferred-momentum scattering events were considered 
important if the total collision rate per atom far exceeded the value of $\Gamma ^\mathrm{damp}$. 
This corresponds, in the limit 
\begin{equation} 
k_0\zeta _\mathrm{h} \gg  1     ,              \label{w.50c} 
\end{equation}  
to temperatures 
$T\gtrsim g_\mathrm{1D}n_\mathrm{1D} (k_0\zeta _{h})^{1/2}$. We show, that this condition is not sufficient for 
experimental observation of the effect of these collisions, and require Eq. (\ref{w.49}) instead. The  shift  of the mean momentum 
of fast particles becomes significant at  
\begin{equation} 
T \gtrsim g_\mathrm{1D}n_\mathrm{1D} (k_0\zeta _{h})^{4/3}(\Delta k\zeta _{h})^{2/3}=\frac {\hbar ^2 k_0^{4/3}\Delta k^{2/3}}m, 
\label{w.50a}
\end{equation} 
and large diffusive spreading of the momentum distribution of fast particles requires 
\begin{equation} 
T \gtrsim g_\mathrm{1D}n_\mathrm{1D} (k_0\zeta _{h})^{6/5}(\Delta k\zeta _{h})^{4/5}=\frac {\hbar ^2 k_0^{6/5}\Delta k^{4/5}}m.  
\label{w.50b}
\end{equation}
Note that Eq. (\ref{w.50c}) does not hold for standard two-photon Bragg spectroscopy \cite{Bragg1}, which can provide 
$k_0\zeta _\mathrm{h}\sim 5$. In this case, a rough estimation yields, instead of Eq.(\ref{w.50a}, \ref{w.50b}), $T\gtrsim 
g_\mathrm{1D}n_\mathrm{1D}$ (cf. Fig.~\ref{F4}). 

\section{Discussion and conclusions} 
\label{sec.5} 
Eq. (\ref{w.49}) is not the only restriction that applies in a real experiment. In practice, we always need to take into account 
the finiteness of the system. The longitudinal trapping determines the finite size of the ultracold atomic cloud \cite{radius}. 
This size can be associated, by order of magnitude, with the parameter $\ell $ introduced in Sec. \ref{sec:3.1}. The longitudinal trapping 
potential itself lifts, in principle, the integrability of the Lieb-Liniger model, however, we expect this non-integrability effect to be 
small \cite{ti11}. More important is anharmonicity (of the longitudinal trap itself or induced by the atomic mean-field interactions) 
that induces dephasing and thus limits the time of observation of the integrable dynamics to few oscillation periods \cite{WeissNC}. 
The most desirable case  takes place if the bosonically-amplified three-body elastic processes manifest themselves after a single  
passage of fast atoms through the cloud, i.e. on the time scale $\tau _0 = m\ell /(\hbar k_0)$. 
The Bragg pulse that produces the fast atoms should be also much shorter than $\tau _0$. In contrast to the Fourier-limited width of the 
excitation spectrum in the frequency domain, the momentum distribution width $\Delta k \approx mT/(2\hbar ^2{n}_\mathrm{1D})$ 
of the fast particles is determined by thermal fluctuations 
of the phase of the quasicondensate \cite{Aspect1}. Regarding a longitudinally trapped quasicondensate, we will use the notation 
${n}_\mathrm{1D} $ for its average linear density. Although the momentum distribution is Lorentzian \cite{Aspect1,Aspect2}, we still use  
Eq. (\ref{w.48}), which is derived in the case of a Gaussian, for rough estimations. We assume that, after the Bragg pulse, fast atoms are 
left to interact with the rest of the quasicondensate for the time $t_\mathrm{int}\sim  \tau _0$. Then the trap is suddenly switched off, 
and the atomic cloud expands ballistically. The atomic position distribution in the asymptotic  regime of long time of flight 
corresponds to the in-trap momentum distribution \cite{Bragg1}. We assume that 
the influence of the bosonically amplified three-body 
collisions with small momentum transfer is unambiguously detectable  
if the relative shift of the momentum of fast stoms due to three-body elastic scattering is of about 50\%, i.e. 
$At_\mathrm{int} \approx  0.5 \, k_0$. To be definite, consider an atomic cloud of size $\ell \approx 100~\mu $m. Initial wave number 
$k_0=0.7 \times 10^5 \; \mathrm{cm}^{-1}$ of the fast particles corresponds to $\tau _0 \approx 20 $ ms. Taking $t_\mathrm{int} = 10$ ms, 
we find that the bosonically-amplified three-body scattering is detectable at $T>110$ nK for $n_\mathrm{1D}=80~\mu \mathrm{m}^{-1}$ and 
$T>140$ nK for $n_\mathrm{1D}=60~\mu \mathrm{m}^{-1}$. The linear density  $n_\mathrm{1D}=40~\mu \mathrm{m}^{-1}$ is too small, and the 
slowing down of fast particles can not be  seen in the temperature range corresponding to the 1D regime. Note for all these density values 
$\Gamma ^\mathrm{damp}t_\mathrm{int} <1$, i.e. three-body scattering into empty modes (with lage momentum transfer) is too weak to be 
detected in a singe passage of fast atoms through the quasicondensate. 

An alternative (to Bragg spectroscopy) 
way of producing fast atoms in twin-atom beams by external driving of the radial excitations of ultracold atoms 
in an elongated trap has been recently demonstrated by B\"ucker \textit{et al.} \cite{Bue}. 

Finally, we discuss the relation between the theory of ultracold atomic systems in the quasi-1D regime 
and the quantum Newton's cradle experiment \cite{WeissNC} that 
has brought a new attention to the problem of the integrability breakdown.  
Thermalization has not been observed in that experiment, and only lower limits to the thermalization time 
have been obtained for different 
(strong and intermediate) interaction strengths \cite{WeissNC}. This fact is in a quantitative agreement with the estimations 
of the thermalization rate \cite{Glazman} (see also Ref. \cite{Mazets2}) taking into account the effect of the 
atomic correlations [the zero-distance two-body correlation function $g_2(0)<1$] in the case of strong 
repulsive interactions. The damping 
time $1/[\gamma _0 g_2(0)]$ occurs  
to be too long to be measured. Also a small size of the colliding atomic clouds  puts a relatively high infrared momentum 
cut-off ($2\pi /\ell >0.1~\mu \mathrm{m}^{-1}$) that precludes scattering events with small momentum transfer. 
As we mentioned before, non-integrability caused by the longitudinal harmonic confinement is also too weak to be detected 
\cite{ti11}. Another possible mechanism of the non-integrability, associated with the tunnel coupling between adjacent 
waveguides in a 2D optical lattice, can be relevant only for optical lattices much weaker than that    
applied in Ref. \cite{WeissNC}. To observe thermalization of bosonic atoms in a quasi-1D waveguide 
and, in particular, to test the theory developed in the present paper, one 
has to work at larger atom numbers that provide both higher atomic linear densities 
and longer system sizes than in the quantum Newton's cradle experiment \cite{WeissNC}.

To summarize, 
in this paper we analyzed effective three-body collisions (mediated by virtual excitations of the radial degrees of freedom) in a 1D 
quasicondensate of ultracold bosonic atoms. We calculated the matrix element for the decay of a single elementary excitations into 
three elementary excitations. We stress that the obtained expression is non-divergent at small momentum transfer. We derived a  
kinetic equation governing the damping of fast particles in a quasicondensate and its Fokker-Planck limit that accounts for 
scattering into thermally populated modes with small momenta. We demonstrate that the latter process can be observed experimentally.

The author thanks J. Armijo, I. Bouchoule, L. I. Glazman, H. Grosse, 
S. Manz, J. Schmiedmayer, and J. Yngvason for helpful discussions. 
This work is supported by the FWF (project P22590-N16).


\begin{thebibliography}{99}
\bibitem{Aspect2} The shape and the width of the momentum distribution of atoms decoupled by the Bragg pulse 
\cite{Aspect1} are in a good agreement with their estimation from the Fourier transform of the first-order 
spatial correlation function of a weakly-interacting quasicondensate \cite{Popov,MC}.
\end{thebibliography}
\end{document}